\documentclass[10pt,aps,prb,groupedaddress,showpacs,twocolumn,showkeys]{revtex4-1}

\usepackage{amsmath}
\usepackage{amssymb}
\usepackage{amsfonts}
\usepackage{color}
\usepackage[pdftex]{graphicx}
\usepackage{epstopdf}
\usepackage{bm}
\usepackage{float}
\usepackage[justification=justified,singlelinecheck=false]{caption}
\usepackage{subcaption}

\begin{document}
\title{Optimal interactions of light with magnetic and electric resonant particles}
\author{R\'{e}mi Colom}
\affiliation{Aix-Marseille Universit\'e, Centrale Marseille, CNRS, Institut Fresnel, UMR 7249, Campus de St. J\'er\^ome, 13397 Marseille, France}
\author{Alexis Devilez}
\affiliation{Aix-Marseille Universit\'e, Centrale Marseille, CNRS, Institut Fresnel, UMR 7249, Campus de St. J\'er\^ome, 13397 Marseille, France}
\author{Nicolas Bonod}
\affiliation{Aix-Marseille Universit\'e, Centrale Marseille, CNRS, Institut Fresnel, UMR 7249, Campus de St. J\'er\^ome, 13397 Marseille, France}
\author{Brian Stout}
\affiliation{Aix-Marseille Universit\'e, Centrale Marseille, CNRS, Institut Fresnel, UMR 7249, Campus de St. J\'er\^ome, 13397 Marseille, France}
\email{brian.stout@fresnel.fr}

\begin{abstract}
This work studies the limits of far and near-field electromagnetic response of sub-wavelength scatterers, like the \emph{unitary limit} and 
of lossless scatterers, and the \emph{ideal absorption limit} of lossy particles. These  limit behaviors are described in terms 
of  analytic formulas that approximate finite size effects while rigorously including radiative corrections. This analysis predicts the \emph{electric} and/or \emph{magnetic} limit responses of both metallic and dielectric nanoparticles while quantitatively describing near-field enhancements.
\end{abstract}

\maketitle
\section{Introduction}
Photonic resonances in subwavelength dielectric or metallic scatterers have generated keen interest on
account of their ability to induce strong light-matter interactions near subwavelength particles \cite{Novotny2006,Enoch2012,Agio2013}. Optimizing the resonant interaction between light and such particles appears to be of fundamental importance to improve the efficiency of light scattering and increase the near field enhancements \cite{Atwater2010,Novotny11,Baffou13}. The light-particle interactions are usually quantified by calculating the absorption and scattering cross sections \cite{Bohren1983,Mishchenko2002}. The question addressed in this work is how to reach their theoretical limits in order to maximize light scattering or absorption by metallic or dielectric photonic resonators, and how to quantify the accompanying near-field enhancements.  

The T-matrix formalism\cite{Khleb13} has long proven quite useful in describing light scattering by particles since it provides a \emph{complete} scattering solution in a relatively intuitive manner. 
However, recent literature in optics has emphasized that alternate scattering formulations,
like the $S$ and $K$ matrices, provide useful complementary descriptions of light-matter interactions that shed additional light on conservation laws and limit behaviors.\cite{Grigoriev2015,LeRu2013}

In this work, we use these alternative formulations to derive approximate
formulas which can describe the resonant and off-resonant response of small spheres of any size and composition.
Although the studied optical response limits also apply to particles of arbitrary shape, spherical symmetry  
is ideal for analytically defining the limit behaviors. More precisely, we use Laurent expansion of the inverse $K$-matrix, we obtain highly accurate energy-conserving approximations to the electromagnetic response of small particles.

These accurate approximations are used to define the Unitary Limit (UL) and Ideal Absorption (IA) conditions. 
The limits are first investigated in the case of plasmonic nanoparticles, but a particular interest is then devoted to the
magnetic modes hosted by dielectric particles. Magnetic modes are very promising to enhance light matter interaction
and we derive the analytic formulas of the dielectric permittivities that optimize light matter interaction \textit{via} the
magnetic mode excitation \cite{Rolly2012,Shcherbakov2014,Boudarham2014,Bakker2015,Zambrana2015,Makarov2015}. We also give formulas that can accurately predict the near-field enhancements around the particles. 

\section{Electromagnetic response of subwavelength sized particles}
\label{Kmatrix}

Let us consider a scatterer characterized by a permittivity $\varepsilon_{s}$ and permeability $\mu_{s}$,
placed in a background medium of permittivity $\varepsilon_{b}$, and permeability $\mu_{b}$ 
(index $N_b=\sqrt{\varepsilon_{b}\mu_{b}}$).
The scattering by a homogeneous spherically symmetric scatterer is completely characterized by its multipolar
Mie coefficients which are opposite in sign to the T-matrix coefficients detailed in Appendix \ref{Theory}, 
{\it} $a_{n} =-T_{n}^{(e)}$ and $b_{n} =-T_{n}^{(h)}$ (where superscripted $(e)$ and $(h)$ designate electric and magnetic modes respectively). 
The relations of Eq.(\ref{TK}) of Appendix \ref{Theory} then allow one to write the Mie coefficients
in terms of the inverse $K$ matrix elements\cite{LeRu2013} ( see appendix \ref{Theory} ):
\begin{subequations}
\begin{eqnarray}
a_{n}^{-1} &=&-i(K_{n}^{(e)})^{-1}+1  \label{a1}\\
b_{n}^{-1} &=&-i(K_{n}^{(h)})^{-1}+1 \label{b1}
\end{eqnarray}
\label{Kinvexp}
\end{subequations}

There is a distinct advantage in writing the Mie coefficients in this manner since as shown in Appendix \ref{Theory}, the reaction matrix elements, $K_{n}$, must be real valued for lossless scatterers with no intrinsic loss. Consequently, any
real-valued approximation to $K_{n}$ in Eq.(\ref{K1}) will preserve the energy conservation relations for the Mie coefficients of \emph{lossless}
scatterers ({\it i.e.} ${\rm Re}\{a_n\}=\vert a_n\vert^2$ and ${\rm Re}\{b_n\}=\vert b_n\vert^2$   for lossless scatterers). 

The relations of Eq.(\ref{Kinvexp}) are a multipole generalization of the well known 
\emph{energy conserving} representation of the electric dipole polarizability, $\alpha_{e}$, where $\alpha_{e}$ 
embodies the linear relationship between the excitation field and the object's induced electric dipole moment, $\mathbf{p} = \epsilon_0\varepsilon_{b}
\alpha_{e}\mathbf{E}_{\rm exc}$. Energy conserving
approximations to the frequency dependent polarizability, $\alpha_{e}$, have long been known to take the form:
\cite{Grigoriev2015,Colas2008}
\begin{equation}
{\alpha}_{e}^{-1}=\alpha_{\rm n.r.}^{-1} - i\frac{k^{3}}{6\pi} \label{polconsv}
\end{equation}
where $k=2\pi/\lambda$ is the in-medium wavenumber. The term $\alpha_{\rm n.r.}$ is a ``non-radiative'' 
polarizability, often approximated by its electrostatic value even though finite size corrections have repeatedly been 
proposed.\cite{Grigoriev2015,Moroz:09,Colas2008,ColasdesFrancs:08,Moroz:09,LeRu2013,Albaladejo10} 
Recalling the relation between the dimensionless Mie coefficients and polarizability\cite{Stout11},  
$\alpha_{e}= i\frac{6\pi}{k^{3}} a_1$, and defining  the ``non-radiative'' polarizability
in terms of the reaction matrix as $\alpha_{\rm n.r.}\equiv -6\pi K_{1}^{(e)}/k^3$, Eq.(\ref{polconsv}) becomes synonymous with the electric dipole case of Eq.(\ref{Kinvexp}) up to an overall multiplicative factor.

The Laurent series development of the inverse reaction matrix elements, $K_{n}$, in powers of $x=kR$ is given in Appendix \ref{6ord}. Restricting attention to the dipole term, $n=1$, for small scatterers yields:
\begin{subequations}
\begin{align}
[K_{1}^{(e)}]^{-1} &  =-\frac{3(\overline{\varepsilon}_{s}+2)}{2(kR)^{3}
(\overline{\varepsilon}_{s}-1)}\left(  1-\frac{3(kR)^{2}(\overline{\varepsilon}_{s}-2)}
{5(\overline{\varepsilon_{s}}+2)} \right. \notag \\ &\qquad \left. -
\frac{3(kR)^{4}(\overline{\varepsilon}_{s}^{2}-24\overline{\varepsilon}_{s}+16)}{350(\overline{\varepsilon}_{s}+2)}
\right)  \label{K1e}  \\
[K_{1}^{(h)}]^{-1} &  =-\frac{45}{(kR)^{5}(\overline
{\varepsilon}_{s}-1)}\left(  1-\frac{(kR)^{2}(2\overline{\varepsilon}_{s}-5)}{21} \right. \notag \\
& \qquad\left. -\frac{(kR)^{4}(\overline{\varepsilon}_{s}^{2}+100\overline
{\varepsilon}_{s}-125)}{2205}\right) \label{K1h}
\end{align}
\label{K1}
\end{subequations}
The fourth order, $x^4$ size corrections, of eqs.(\ref{K1}), will usually suffice,  
but the order $x^6$ size corrections in the parenthesis are given in Appendix \ref{6ord}.

\begin{figure}[H]
\includegraphics[width=1\columnwidth]{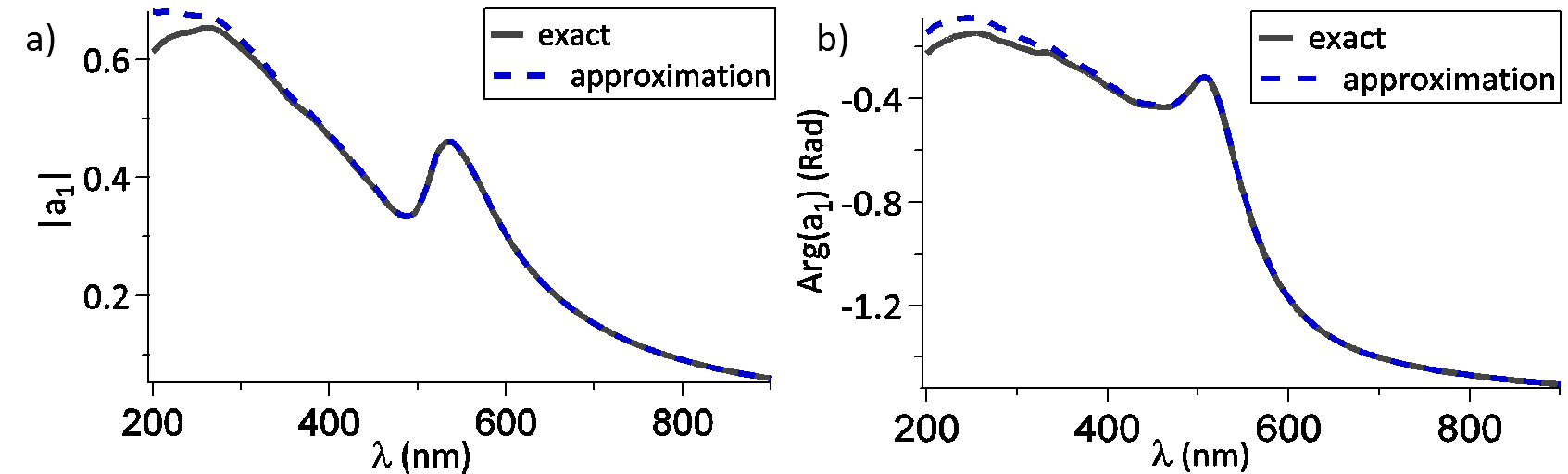}
\caption{(Colour online) Approximate values for the electric dipole Mie coefficient, $a_{1}$, (amplitude and phase) compared with the exact values (solid black) ; $R=60$nm gold sphere.}
\label{a1_gold_sphere}
\end{figure}

Long wavelength approximations of the electric mode Mie coefficients, like that found by inserting Eq.(\ref{K1e}) into
 Eq.(\ref{a1}), have already been shown to accurately describe the electric response of small metallic spheres\cite{LeRu2013} as illustrated in Fig.(\ref{a1_gold_sphere}) where we compare exact and approximate $a_{1}$ coefficients of a $R=60$nm radius gold sphere.\cite{Johnson}			
\begin{figure}[H]
\centering\includegraphics[width=1\columnwidth]{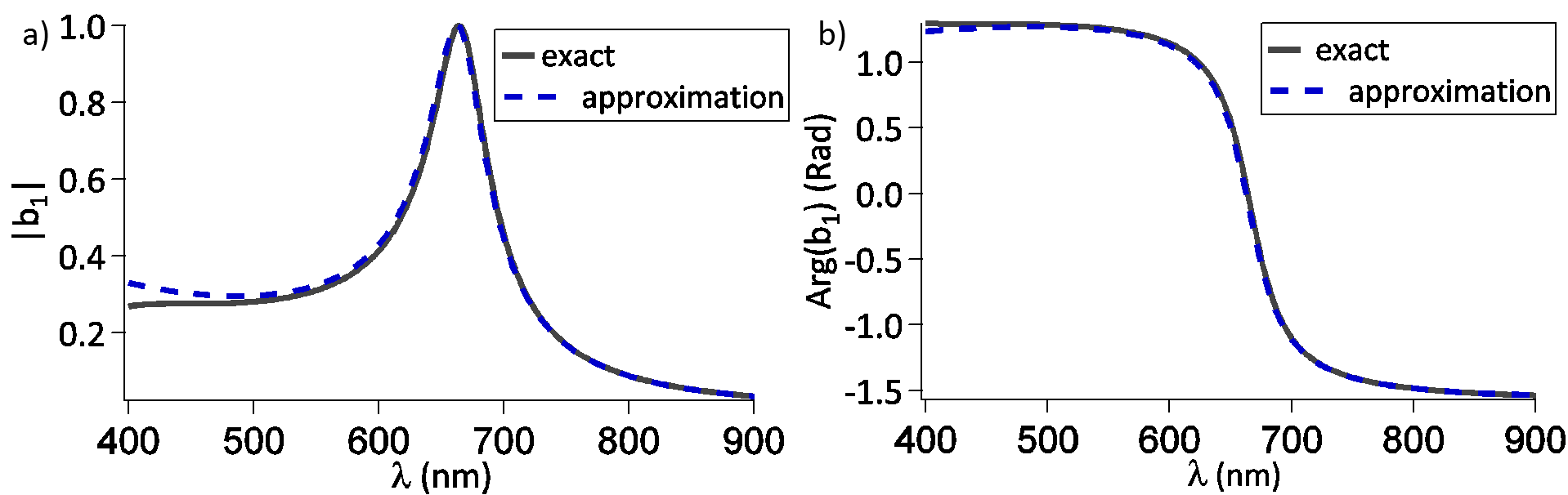}
\caption{(Colour online) Approximate values for the magnetic dipole Mie coefficient, $b_{1}$, (amplitude and phase) compared with the exact values (solid black) ; $R=80$nm, $\varepsilon=16$.}
\label{b1_dielectric_sphere}
\end{figure}

It is much less appreciated however, that the analogous procedure for magnetic modes of inserting 
Eq.(\ref{K1h}) into Eq.(\ref{b1}) produces good approximations to magnetic dipole resonances in high
index dielectric spheres. The results displayed in Fig.(\ref{b1_dielectric_sphere}) illustrate the accuracy of this method to describe the magnetic dipole coefficient of a $R=80$nm, $\varepsilon=16$ dielectric sphere (despite the fact that that the quasi-static magnetic polarizability is zero due to the absence of permeability contrasts).


\section{Optimal light-particle interactions}
\label{ULIAcond}

\subsection{Limit response conditions}
\label{Limcond}

The total cross sections of an arbitrary spherically symmetric scatterer are the sum of the contributions from all the multipolar modes:
\begin{equation}
\sigma  =\sum\limits_{n=1}^{\infty }\left\{ \sigma _{n}^{(e)}+\sigma_{n}^{(h)}
\right\}
\end{equation}
where $\sigma$ can be either the scattering, ($\sigma _{\rm s}$), extinction ($\sigma _{\rm e}$), 
or absorption ($\sigma _{\rm a}$), cross sections and the $n^{th}$ multipolar mode 
contributions respectively denoted $\sigma _{n,\rm s}^{(q)}$, $\sigma _{n,\rm e}^{(q)}$, 
$\sigma _{n,\rm a}^{(q)}$, with $(q)=(e)$ or $(h)$ for (electric or magnetic modes) .

Expressing the modal contributions to the cross sections, $\sigma _{\rm s}$, $\sigma _{\rm e}$, 
and $\sigma _{\rm a}$, in terms of the $S$ matrix is particularly convenient for determining optical response limits:\cite{Grigoriev2015}
\begin{align}
\begin{split}
\sigma _{n,\rm e}^{(q)}&=\frac{(2n+1)\lambda ^{2}}{4\pi }
{\rm Re}\left\{ 1- S_{n}^{(q)} \right\}  \\
\sigma_{n,\rm s}^{(q)} & = \frac{(2n+1)\lambda ^{2}}{8\pi } \left\vert 1-S_{n}^{(q)} \right\vert^2\ ,
 \\
\sigma _{n,\rm a}^{(q)} &=\frac{(2n+1)\lambda ^{2}}{8\pi }
\left( 1-\left\vert S_{n}^{(q)}\right\vert ^{2}\right) 
\end{split}
\label{Scross}
\end{align}
where $\lambda=\lambda_0/N_b$ is the in-medium wavelength and we recall that $\vert S_{1}^{(q)} \vert \leq 1$ for passive media 
( $\vert S_{1}^{(q)} \vert = 1$ for lossless media). The $(2n+1)$ factor in Eq.(\ref{Scross}) results from
the $2n+1$ degeneracy of the projection quantum numbers for each angular momentum number, $n$.

We henceforth adopt the usual definition for the unitary limit (UL) as occurring whenever the contribution to the scattering cross section
of at least one mode reaches its upper bound while Ideal Absorption (IA) analogously occurs whenever the contribution to the
absorption cross section of one mode attains its upper bound.\cite{Grigoriev2015,Alucloak14,Tret14}.

Therefore, Eqs.(\ref{Scross}) show that the UL and IA conditions (\emph{of a given mode at a given wavelength}) can be expressed in terms of the $S$ matrix  coefficients as:
\begin{equation}\label{IA_condition_S}
S_{n, \rm UL}^{(q)}=-1 \quad \ , \quad S_{n, \rm IA}^{(q)}=0
\end{equation}
The cross sections bounds associated with UL or IA in an $n^{th}$ order mode are respectively:
\begin{equation} \label{IA_sigma}
\begin{tabular}{cc}
Unitary limit (UL) \quad  & \quad Ideal absorption (IA) \\
\qquad  & \qquad \\
$\sigma_{n,\rm s}^{(q)}=\sigma_{n,\rm e}^{(q)}=\frac{(2n+1)\lambda^{2}}{2\pi}$ \ \ , & \ 
$\sigma_{n,\rm a}^{(q)}=\sigma_{n,\rm s}^{(q)}=\frac{(2n+1)\lambda^{2}}{8\pi}$
\end{tabular}
\end{equation}


The UL and IA conditions in terms of reaction matrix coefficients $K_{n}$ are readily found by respectively inserting the conditions of Eq.(\ref{IA_condition_S}) into the Eq.(\ref{Cayley}) Cayley transform relations:
\begin{equation}
[K_{n,{\rm UL}}^{(q)}]^{-1}=0 \quad , \quad [K_{n,{\rm IA}}^{(q)}]^{-1}=i
\label{KULIA}
\end{equation}

\begin{figure}[H]
\includegraphics[width=1\columnwidth]{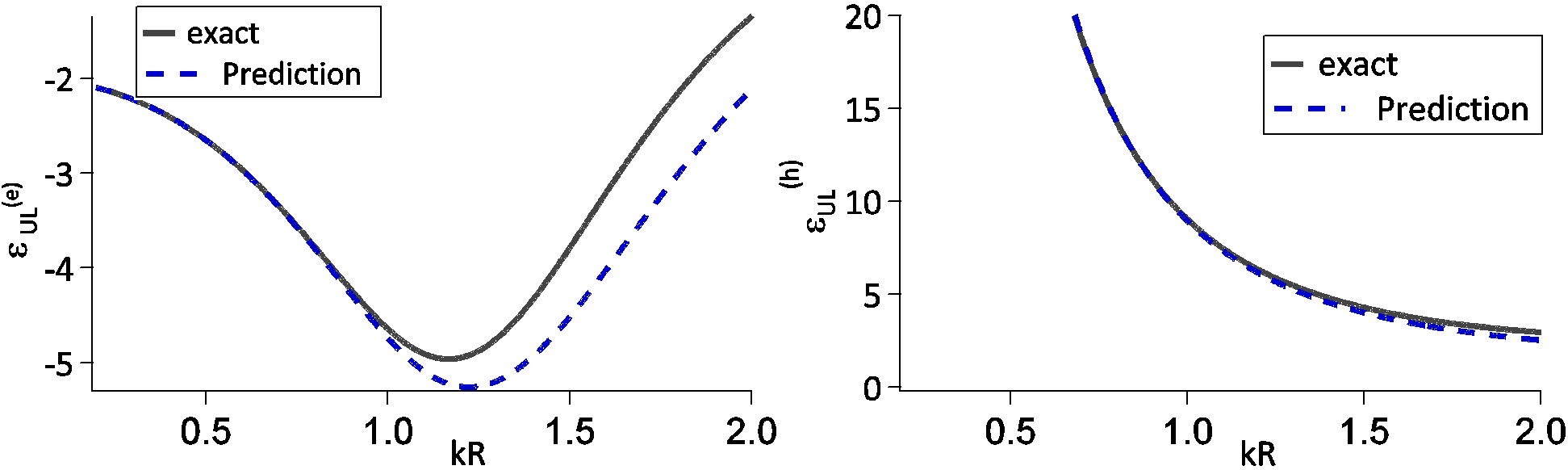}
\caption{(Colour online) Values of permittivity required to reach the dipole UL in the electric (a) and magnetic (b) modes as a function of $kR$. Exact predictions (full black lines) and approximate predictions  (dashed blue). 
} \label{unitaryfig}
\end{figure}

In Fig.(\ref{unitaryfig}), we compare the exact solutions of the unitary limit of Eq.(\ref{KULIA}) in terms of permittivity as a function of the particle size parameter, $kR$, with the algebraically obtained predictions employing the approximations of Eq.(\ref{K1}). One remarks that the UL permittivities are real valued as required by unitarity and inspection of  Eq.(\ref{Scross}).
\begin{figure}[H]
\centering\includegraphics[width=1\columnwidth]{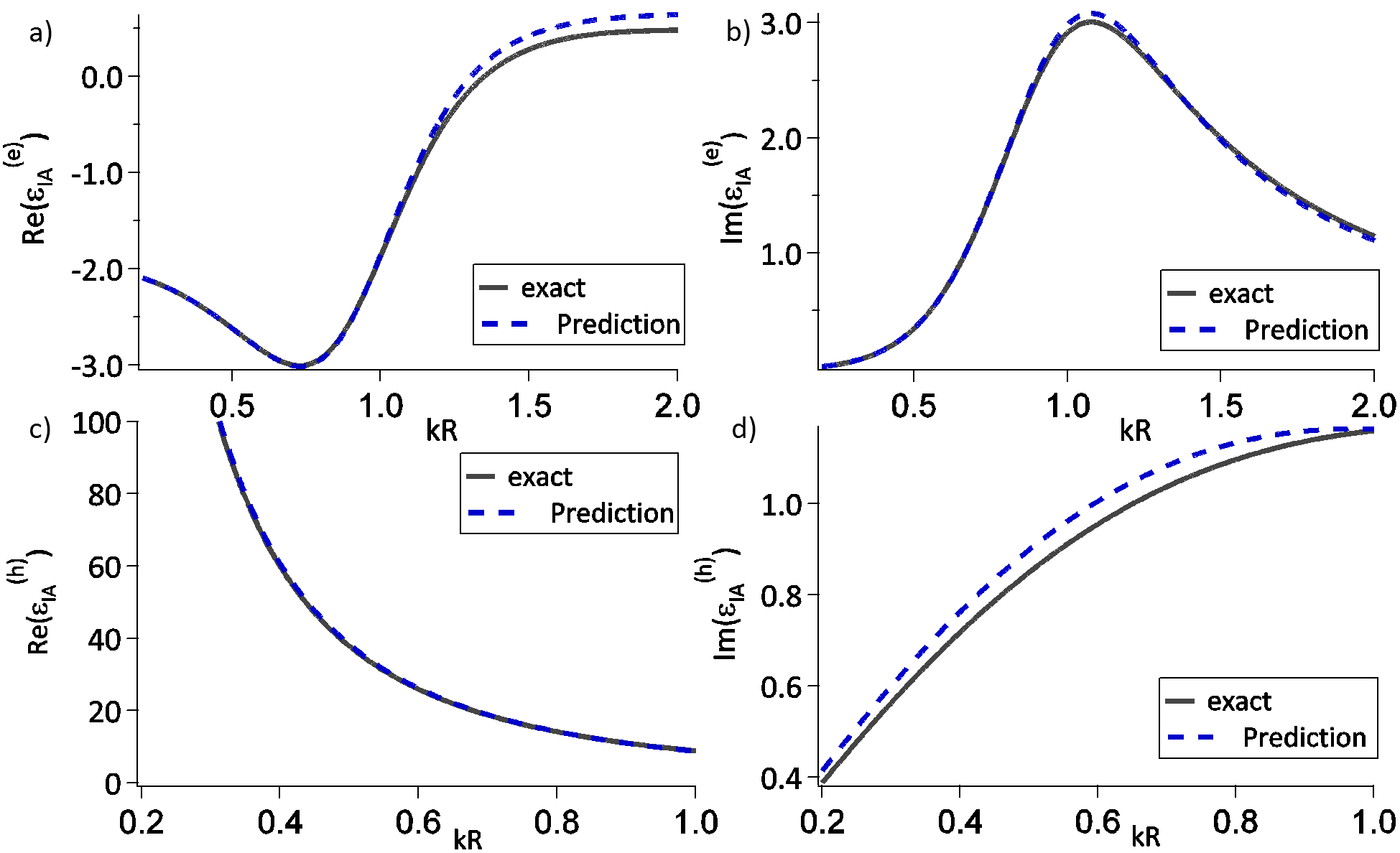}
\caption{(Colour online) Real (a,c) and imaginary (b,d) parts of the dielectric permittivity satisfying IA in the electric (a,b) and magnetic (c,d) dipole modes. Exact predictions (full black lines) and approximate predictions (dashed blue lines).}
\label{Re_Im_eps_IA_dip_elec}
\end{figure}

Approximate predictions of IA can likewise be obtained by solving the IA condition in Eq.(\ref{KULIA}) with the
approximate expression of $K_{1}^{(q)}$ given in Eq.(\ref{K1}). These are compared with exact calculations in Fig.(\ref{Re_Im_eps_IA_dip_elec}). Approximate expressions are even more useful here since exact IA solutions require solving a complex transcendental equation.

One remarks that the approximate predictions for both UL and IA in Figs.(\ref{unitaryfig}) and (\ref{Re_Im_eps_IA_dip_elec}) are in good agreement for small size parameters ($kR < 1$) and remain close even for larger $kR$. 
In practice, the electric mode limit responses are most readily attainable
with materials possessing plasmonic responses, like gold or silver, while for magnetic modes, high-index low-loss  
materials like silicon, $\varepsilon_{\rm Si} \sim 14$ are required.

\subsection{Near and far field spectrum}
We saw above that the use of Eqs.(\ref{Kinvexp}), (\ref{K1}), and (\ref{Scross}) allows a 
rapid determination of the optimal far field responses of small particles, but resonant response is also
of interest due to the near-field enhancements it induces. It has recently been pointed out in recent studies however, 
that there is a red shift of the optimal near field enhancements with respect to the cross section maxima.\cite{Moreno13,Menzel2014,Yuffa2015} 

We derived approximate and \emph{exact} formulas in Eqs.(\ref{Ienh}) and (\ref{Intexact}) respectively in order 
to quantify these near-field spectral shifts for \emph{both} electric and magnetic field enhancements. 
Our approximate expression for the electric field enhancement factor $\left< I_{\mathrm{enh}}^{(e)}\right>$, given in Eq.(\ref{Ieenh})
below, is quite similar to a formula derived by Yuffa {\it et al.},\cite{Yuffa2015} but those authors used somewhat different definitions of field enhancements (apparently due to the fact that they
looked at scattered fields rather than the total fields considered here).

Angle-averaged local electric and magnetic field intensity enhancement factors, 
$\left< I_{\mathrm{enh}}^{(e)}\right>$,  and $\left< I_{\mathrm{enh}}^{(h)}\right>$ are
functions of the normalized distance to the particle center, $\eta\equiv kr$, and can be defined as:
\begin{subequations}
\begin{align}
\left< I_{\mathrm{enh}}^{(e)}\right> & \equiv\frac{\int d\Omega\left\Vert
\mathbf{E}_{\mathrm{tot}}(  \eta \widehat{\mathbf{r}}) \right\Vert ^{2}}
{ 4\pi \left\Vert\mathbf{E}_{\mathrm{exc}}\left(  \mathbf{0}\right)  \right\Vert^{2}} \label{Ieenh}\\
&  \simeq 1+ \sum_{n=1}^{\infty} g_{n}^{(1)} (\eta )
\left\vert b_{n} \right\vert^{2}
  + g_{n}^{(2)} (\eta ) \left\vert a_{n}\right\vert ^{2} \nonumber \\
\left< I_{\mathrm{enh}}^{(h)}\right> & \equiv\frac{\int d\Omega\left\Vert
\mathbf{H}_{\mathrm{tot}}\left(  \eta \widehat{\mathbf{r}}\right) \right\Vert ^{2}}
{ 4\pi \left\Vert\mathbf{H}_{\mathrm{exc}}\left(  \mathbf{0}\right)  \right\Vert^{2}} \label{Ihenh}\\
&  \simeq 1+ \sum_{n=1}^{\infty} g_{n}^{(1)} (\eta ) 
\left\vert a_{n}\right\vert ^{2}
  + g_{n}^{(2)} (\eta) \left\vert  b_{n}\right\vert ^{2}\nonumber 
\end{align}
\label{Ienh}
\end{subequations} 
\!\!\!where $\mathbf{E}_{\mathrm{tot}}$ and $\mathbf{H}_{\mathrm{tot}}$ are respectively  
the \emph{total} electric and magnetic fields outside the particle.
The functions $g_{n}^{(1)} (\eta)$ and $g_{n}^{(2)} (\eta )$ of Eq.(\ref{Ienh}) are given by: 
\begin{align}
  \begin{split}
g_{n}^{(1)} (\eta ) & \equiv \frac{ 2n+1}{2} \left\vert h_{n}^{(+)}(\eta )
  \right\vert ^{2} \\
  g_{n}^{(2)} (\eta ) & \equiv \frac{1}{2}\left[ ( n+1)  \left\vert h_{n-1}^{(+)}(\eta ) \right\vert ^{2}
+n\left\vert h_{n+1}^{(+)}(\eta )\right\vert^{2} \right]
\end{split} 
\label{gapp}
\end{align}

The approximation used in deriving Eqs.(\ref{Ienh}) is accurate only as long as the respective electric and magnetic excitation fields, can be approximated by their values at the center of the particle, $\mathbf{E}_{\mathrm{exc}}\left(  \mathbf{0}\right)$ and 
$\mathbf{H}_{\mathrm{exc}}\left(  \mathbf{0}\right)$.
Although this is generally quite accurate at small $kr$ values, its validity can be tested with the exact
expressions for near field enhancements given in appendix \ref{Exact_near}. 

The electric field enhancement formulas in Eq.(\ref{Ienh}) explain why the
maximum of the near field enhancements are generally red-shifted with respect to their cross section maxima. The spherical Hankel functions are rapidly diverging functions in the $kr \rightarrow 0$ limit  (due to existence of evanescent waves near the current sources\cite{Moreno13}), which shifts the near-field maximum to smaller values of $k$ compared to values of $k$ which maximize the amplitude of a Mie coefficient.

\begin{figure}[H]
\begin{centering}\includegraphics[width=0.5\textwidth]{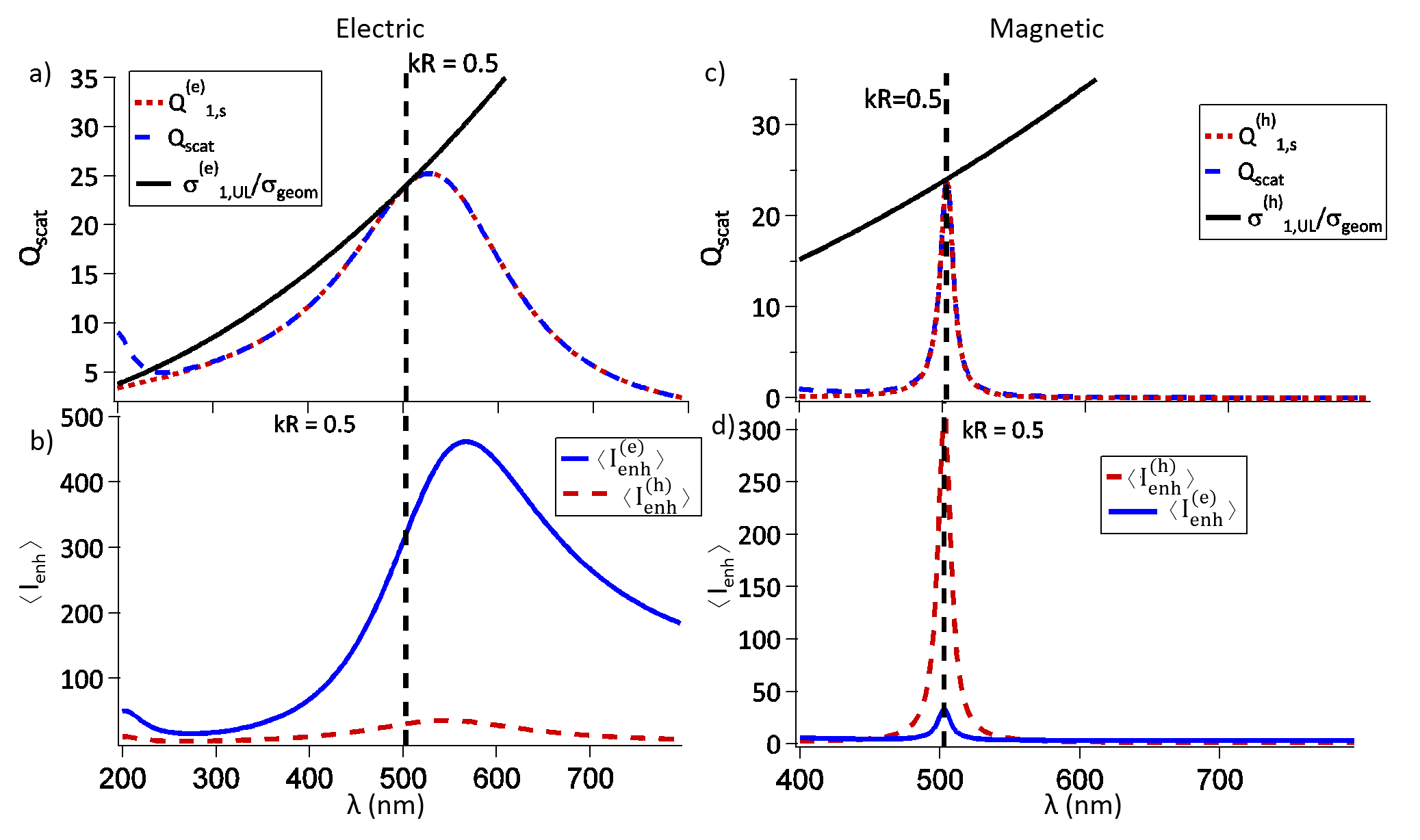}
\end{centering}
\caption{(Colour online) Cross section efficiencies (a,c) and field enhancements (b,d) for $R=40$nm spheres satisfying the dipole UL in electric (a,b) and magnetic (c,d) designed to reach dipole UL at  $kR=0.5$.
}
\label{Q_e_h_UL}
\end{figure}
\begin{figure}[H]
\includegraphics[width=0.5\textwidth]{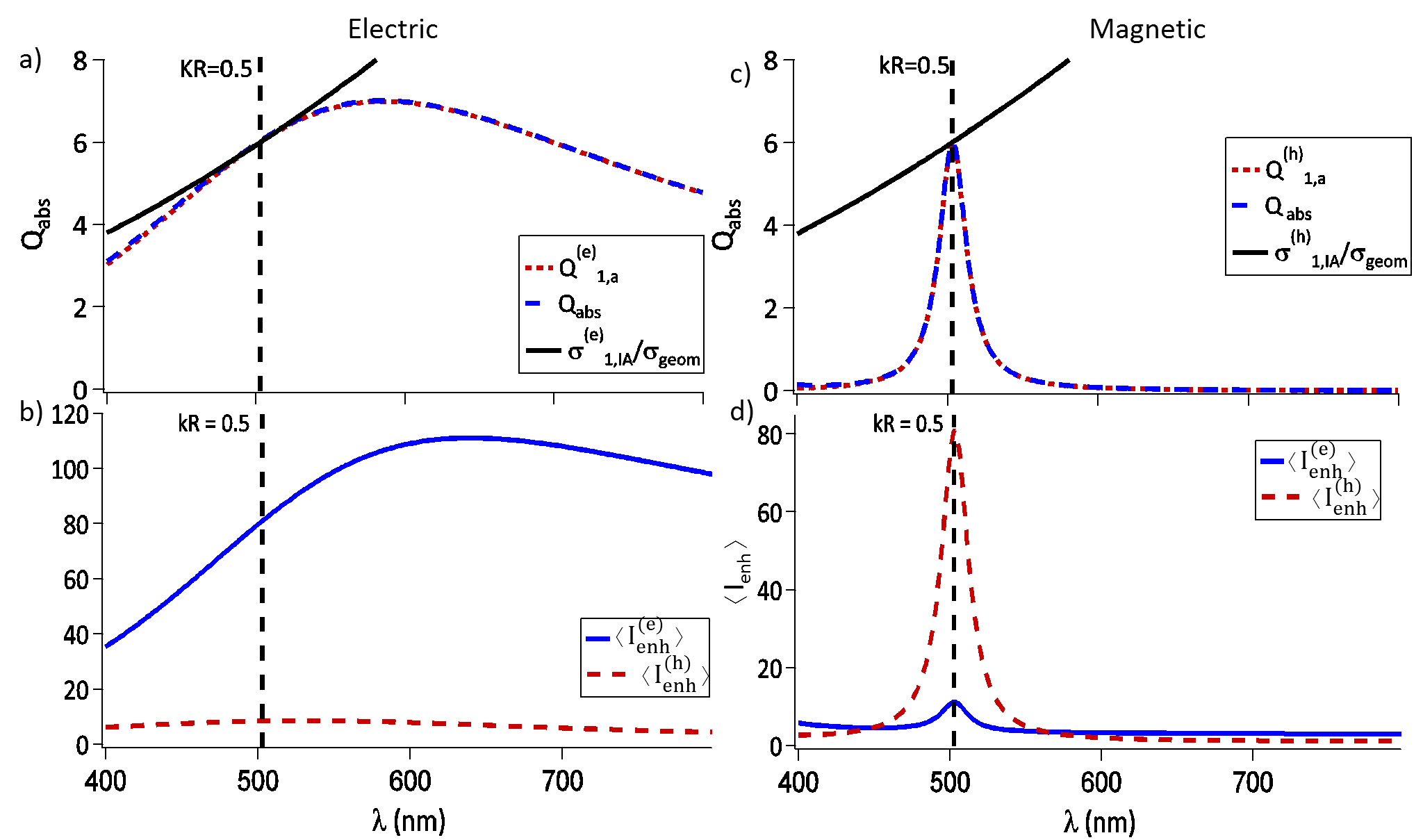} 
\caption{(Colour online) Same plots as in Fig.\ref{Q_e_h_UL} except for plotting absorption efficiencies in (a,c)
in dipole IA particles at $kR=0.5$.}           
\label{Q_abs_IA_dip_elec_mag_kR05}
\end{figure}

In Figs.(\ref{Q_e_h_UL}a,c), we plot the scattering efficiencies, 
$Q_{\rm scat}=\sigma_{\rm scat}/{\sigma_{\rm geom}}$ ($\sigma_{\rm geom}=\pi R^{2}$ 
is the geometrical cross section) , of $R=40$nm radii spheres whose permittivities are chosen so that 
electric (magnetic) dipole 
ULs are respectively reached when $kR=0.5$ ($\overline{\varepsilon}_{UL}^{(e)}= -2.65$,
$\overline{\varepsilon}_{UL}^{(h)}= 37.9$). Total efficiencies are plotted in dashed blue while dipole electric and
magnetic contributions are plotted in dotted red. Unitary limit dipole efficiencies are plotted as solid black
lines as a reference. 
Angle averaged field enhancements,  $\left< I_{\mathrm{enh}}^{(e,h)}\right>$ for both electric and magnetic fields are plotted in Figs.(\ref{Q_e_h_UL}b,d).

We remark  that the cross sections in Fig.(\ref{Q_e_h_UL}a), being weighted by the wavelength squared as seen in 
Eq.(\ref{Scross}), have their maximal cross section red-shifted with respect to the UL frequency.  The maximum field
enhancements $\left< I_{\mathrm{enh}}^{(e,h)}\right>$, are however even more red shifted than the cross sections
according to the arguments presented after Eq.(\ref{gapp}). Both red-shifts are far less pronounced for narrower
resonances like those of the magnetic dipole UL and IA in Figs.(\ref{Q_e_h_UL}c,d) 
and (\ref{Q_abs_IA_dip_elec_mag_kR05}c,d).

Like in the UL case, the spectral behavior of IA  spheres can be studied by plotting the evolution of the absorption
efficiency, $Q_{\rm abs}=\sigma_{\rm abs}/{\sigma_{\rm geom}}$ as shown in 
Fig.(\ref{Q_abs_IA_dip_elec_mag_kR05}a,c), for the dipole electric (a) and magnetic (b) modes of $R=40$nm
spheres designed to reach IA at a size parameter of $kR=0.5$ ( $\overline{\varepsilon}_{\rm IA}^{(e)}= -2.62+0.35i $, $\overline{\varepsilon}_{\rm IA}^{(h)}=37.9+i0.85$).

\section{Algebraic expressions for optimal magnetic light-particle interactions }

\begin{figure}[H]
\centering\includegraphics[width=1\columnwidth]{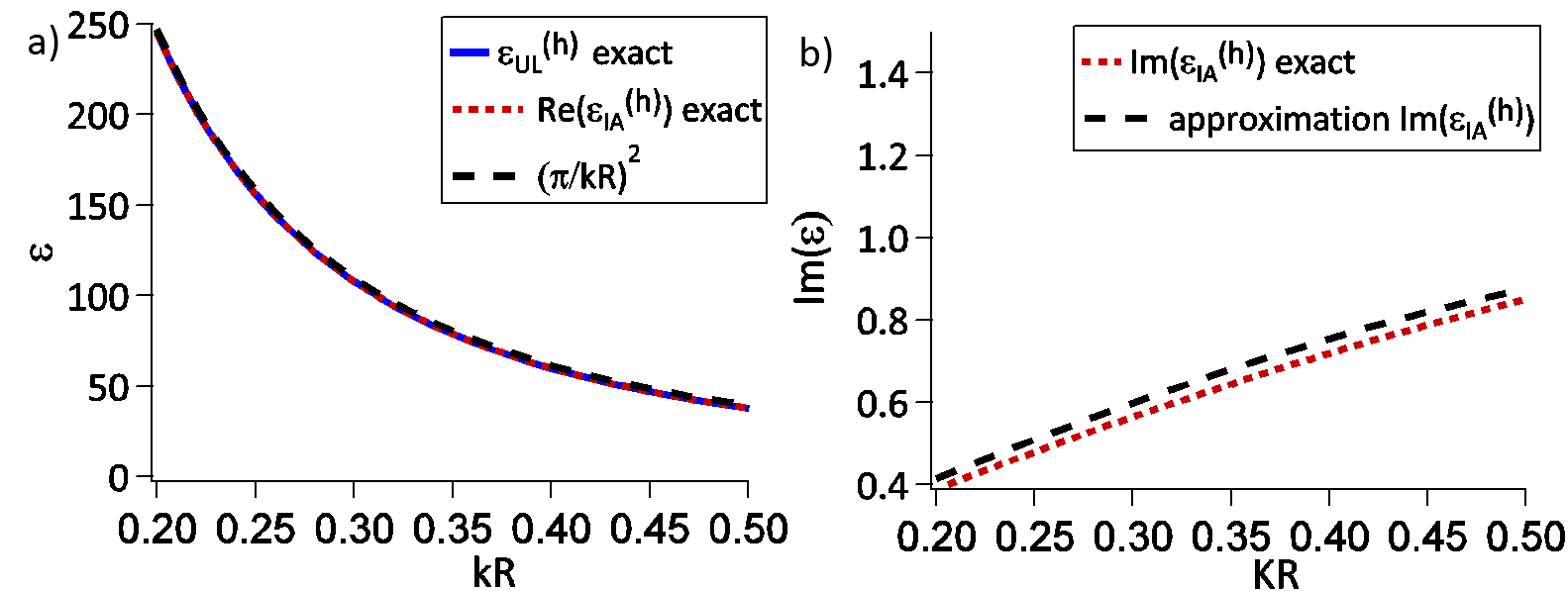}
\caption{(Colour online) Permittivities satisfying magnetic dipole  UL, $\varepsilon _{\rm UL}^{(h)}$, (solid blue) and
satisfying magnetic IA, $\varepsilon _{\rm IA}^{(h)}$ (dotted red) as functions of $kR$: Real parts (a) and
imaginary parts (b). Approximate algebraic expressions of Eqs.(\ref{Re_eps_IA_approx}) and
(\ref{Im_eps_IA_approx}) are plotted in dashed black.}
\label{eps_IA_UL_exact_bis}
\end{figure}

The previous results have shown that the conditions to reach IA and UL in the magnetic dipole mode are in fact very close to one another. From Fig.(\ref{unitaryfig}b) and Fig.(\ref{Re_Im_eps_IA_dip_elec}c), one finds that the permittivity required to reach UL at $kR=0.5$ and the real part of the permittivity necessary to reach IA at that same size are both approximately $\varepsilon\approx 38$. We further illustrate this point by comparing the exact values of 
$\varepsilon_{\rm UL}$ and Re$\{\varepsilon_{\rm IA}\}$ over a range of $kR$ in Fig.(\ref{eps_IA_UL_exact_bis}a).

An explanation of this property is found by examining the limits equations giving the UL and IA conditions (see appendix \ref{Theory} for additional details). From inspection of Eq.(\ref{STK}), one sees that the condition for IA in the magnetic dipole mode is:
\begin{equation}\label{IA_condition}
      S_{1}^{(h)} =0 \iff \varphi _{1}(k_{s}R)=\varphi _{1}^{(-)}(kR)
\end{equation}
where the $\varphi$ functions are defined in Eq.(\ref{phidef}). 
In the small particle limit, $\lim_{x\to 0}h_{n}^{(-)}(x)=-i y_{n}(x)$, and the equation for IA becomes:
\begin{equation}
\text{For } kR\ll 1\ \ S_{1}^{(h)}=0\iff \varphi _{1}(k_{s}R)\backsimeq \varphi _{1}^{(2)}(kR),
\end{equation}
which is identical to the magnetic dipole UL (see Eqs. (\ref{IA_condition_S}), (\ref{KULIA}), and (\ref{STK})):
\begin{equation}
(K_{1}^{(h)})^{-1}=0\iff \varphi _{1}(k_{s}R)=\varphi _{1}^{(2)}(kR).
\label{UL_condition}
\end{equation}

If the previous equations are solved in the $kR \rightarrow 0$ limit, the following simple expressions are found for the IA and UL conditions:
\begin{equation}
\varepsilon _{\rm UL}\backsimeq {\rm Re}(\varepsilon _{\rm IA})\backsimeq \frac{10}
{(kR)^{2}}\backsimeq \left( \frac{\pi }{kR}\right) ^{2}  \label{Re_eps_IA_approx}
\end{equation}
\begin{equation}
{\rm Im}\{\varepsilon _{\text{IA}}\}\backsimeq \frac{49}{2}\left( 1-\sqrt{\frac{5}{6}}
\right) kR-\frac{203}{24\sqrt{30}}(kR)^{3}  
\label{Im_eps_IA_approx}
\end{equation}
This approximate expression is compared with exact calculations in Fig.(\ref{eps_IA_UL_exact_bis}b).

\section{Transition from unitary limit to ideal absorption}

\begin{figure}[H]
        \centering
        \begin{subfigure}[b]{0.25\textwidth}
                \caption{$kR$=0.4}\includegraphics[width=\textwidth]{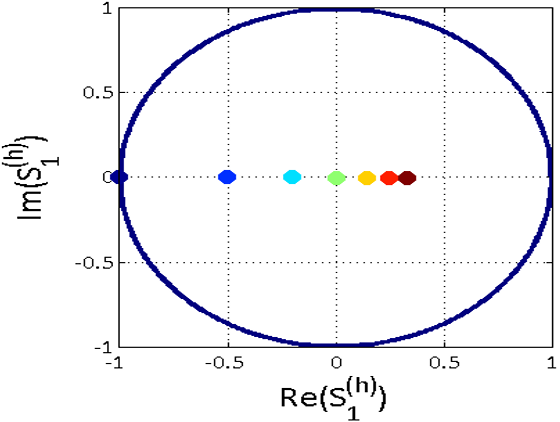}        
				\end{subfigure}%
        \begin{subfigure}[b]{0.25\textwidth}
                \caption{$kR$=0.8}\includegraphics[width=\textwidth]{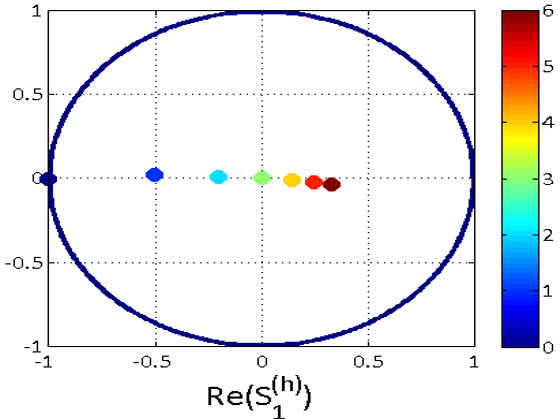}
        \end{subfigure}
				
				\caption{(Colour online) Values of $S_{1}^{(h)}$ for permittivities calculated from Eq.(\ref{eps_UL_plus_2_abs}) for $kR=0.4$ and $kR=0.8$, the color bar indicating the value of $j$ at each point.}
\label{ReS1h_vs_ImS1h_kR__04_08}
\end{figure}

One sees in Fig.(\ref{eps_IA_UL_exact_bis}) that a magnetic dipole UL response can transform into an 
IA response with the appropriate amount of added absorption. This is illustrated in
Fig.(\ref{ReS1h_vs_ImS1h_kR__04_08}) by plotting the values of the complex $S_{1}^{(h)}$ coefficient as the permittivity ranges from the UL permittivity, $\varepsilon _{\rm UL}^{(h)}$, to and beyond the IA permittivity, $\varepsilon_{\rm IA}^{(h)}$, according to the following function for spheres with $kR=0.4$ and $kR=0.8$:
\begin{align}\label{eps_UL_plus_2_abs}
\begin{split}
&\qquad\qquad \varepsilon =\varepsilon _{UL}^{(h)}+j \frac{(\varepsilon _{\rm UL}^{(h)}-\varepsilon _{\rm IA}^{(h)})}{3} \\
& \begin{tabular}{|c|c|}
\hline
$kR=0.4\qquad$ & $kR=0.8$   \\
\hline$\varepsilon _{\rm UL}^{(h)}=59.94\qquad$ & $\varepsilon _{\rm UL}^{(h)}=14.3$ \\
$\varepsilon _{\rm IA}^{(h)}=59.93+i0.72 \quad$ & $\varepsilon _{\rm IA}^{(h)}=14.2+i1.1$\\
\hline
\end{tabular}
\end{split}
\end{align}
where $j$ is an integer between 0 and 6 in each case. The UL value of $S_{1}^{(h)}=-1$ corresponds to $j = 0$, while the 
$S_{1}^{(h)}=0$, IA limit occurs for $j = 3$.
The permittivities, scattering efficiencies, and field enhancements at the surface of the particles for the values of Eq.(\ref{eps_UL_plus_2_abs}) and Fig.(\ref{ReS1h_vs_ImS1h_kR__04_08}) are given in Tables
\ref{kR0p4} and \ref{kR0p8}. A comparison of Tables \ref{kR0p4} and \ref{kR0p8} shows that although the larger
$kR=0.8$ spheres require considerably smaller permittivities, this comes at the expense
of much weaker field enhancements. 
\begin{table}[H]
\begin{tabular}{|c|c|c|c|c|c|c|}
	\hline
  $j$ & $\varepsilon$ & $Q_{1,\rm ext}^{(h)}$ & $Q_{1,\rm scat}^{(h)}$ & $Q_{1,\rm abs}^{(h)}$ & $\left< I_{\mathrm{enh}}^{(h)}\right>$ & $\left< I_{\mathrm{enh}}^{(e)}\right>$  \\
  \hline  0 & 59.94 &  37.5 &  37.5 & 0 & 1168 & 71 \\
  \hline
	1 & $59.94+0.24i$  & 28.1 & 21.1 & 7 & 657 & 42 \\
  \hline
	2 & $59.94+0.48i$  & 22.5 & 13.5 & 9 & 421 & 28 \\
  \hline
	3 & $59.94+0.72i$  & 18.8 & 9.4 & 9.4 & 293 & 20 \\
  \hline
	4 & $59.94+0.96i$  & 16.1  & 6.9 & 9.2 & 215 & 16\\
  \hline
	5 & $59.94+1.2 i$  & 14.1 & 5.3 & 8.8 & 165 & 13\\
  \hline
	6 & $59.94+1.44i$  & 12.5 & 4.2 & 8.3 & 131 & 11\\
   \hline
\end{tabular}
\caption{Cross section and magnetic and electric field enhancement factors for $kR=0.4$ size particles.}
\label{kR0p4}
\end{table} 
\begin{table}[H]
\begin{tabular}{|c|c|c|c|c|c|c|}
	\hline
  $j$ & $\varepsilon$ & $Q_{1,\rm ext}^{(h)}$ & $Q_{1, \rm scat}^{(h)}$ & $Q_{1,\rm abs}^{(h)}$ & $\left< I_{\mathrm{enh}}^{(h)}\right>$ & $\left< I_{\mathrm{enh}}^{(e)}\right>$ \\
  \hline
  0 & 14.3 & 9.4 & 9.4 & 0 & 25 & 10\\
  \hline
	1 & $14.2+0.37i$ & 7  & 5.3 & 1.8 & 14.9 & 7.7 \\
  \hline
	2 & $14.2+0.73i$ & 5.6 & 3.4 & 2.3 & 10.2 & 6.4 \\
  \hline
	3 & $14.2+1.1i$ &  4.7  & 2.3 & 2.4 & 7.6 & 5.7 \\
  \hline
	4 & $14.2+1.5i$ & 4 & 1.7 & 2.3 & 6.1 & 5.3 \\
  \hline
	5 & $14.1+1.8i$ & 3.5 & 1.3 & 2.2 & 5.1 & 5 \\
  \hline
	6 & $14.1+2.2i$  & 3.2  & 1.1 & 2.1 & 4.4 & 4.8\\
 \hline
\end{tabular}
\caption{Cross section and magnetic and electric field enhancement factors for $kR=0.8$ size particles.}
\label{kR0p8}
\end{table}

\section{Conclusion}
In this work, we derived accurate approximate expressions of the particle polarizability, and used these to study unitary limits and ideal absorption limits of the dipolar modes of small particles. We also derived formulas that allow a quantification of the displacement of the near field-maxima with respect to far-field maxima. We applied this approach to both metallic and dielectric nanoparticles and emphasis was placed on magnetic dipolar resonances in high index dielectric particles.  In this latter case, we derived closed expressions for  UL  and IA in the small particle limit. 
This study should help in the design of highly efficient photonic resonators that are of crucial importance to strengthen the light matter interaction at subwavelength scales.

\section{Acknowledgments}
This work has been carried out thanks to the support of the A*MIDEX project (n$^{\rm o}$ ANR-11-IDEX-0001-02) funded by the « Investissements d'Avenir » French Government program, managed by the French National Research Agency (ANR) and
within the context of the France-Australian International Associated Laboratory ``ALPhFA: Associated Laboratory for Photonics". The authors would like to thank Jean-Paul Hugonin and Boris Kuhlmey for fruitful discussions.

\appendix

\section{Multipole Scattering theory}
\label{Theory}

Notations in the literature vary considerably, and we begin this section by reviewing our notation for vector partial waves ( VPW) expansions. Let us consider an arbitrary-shaped scatterer characterized by a permittivity $\varepsilon_{s}$ and permeability $\mu_{s}$
placed in a background medium of permittivity $\varepsilon_{b}$, and permeability $\mu_{b}$.

The Foldy-Lax excitation field of a particle can be developed on the basis of regular (source-free) Vector Partial 
Waves (VPWs), of the  first type traditionally denoted $\mathbf{M}_{n,m}^{(1)}$ (magnetic modes) and 
$\mathbf{N}_{n,m}^{(1)}$ (electric electric), ( $n$ and $m$ being respectively the total and projected 
angular momentum of the VPW {\it i.e.}:
\begin{align}
\mathbf{E}_{\mathrm{exc}}(\mathbf{r)}&=\sum\limits_{n,m}
 \left\{
\mathbf{M}_{n,m}^{(1)}(k\mathbf{r})e_{n,m}^{(h)}+\mathbf{N}_{n,m}^{(1)}(k\mathbf{r})e_{n,m}^{(e)} \right\}  \nonumber \\
&\equiv [\mathbf{M}^{(1)}(k\mathbf{r}),\mathbf{N}^{(1)}(k\mathbf{r})]\binom{e^{(h)}}{e^{(e)}}
\label{E_inc_expansion}
\end{align}
where $(e)$ (respectively $(h)$) superscripts indicate electric (respectively
magnetic) type VPWs. In the second line of Eq.(\ref{E_inc_expansion}), we suppressed the 
summations by adopting a more compact matrix notation where expansion
coefficients of the excitation field are placed in an infinite column matrix and the corresponding VPW
functions, $\mathbf{M}_{n,m}^{(1)}(k\mathbf{r})$, and $\mathbf{N}_{n,m}^{(1)}(k\mathbf{r})$, as elements of 
an infinite row ``matrix''.\cite{Tsang01}

The field scattered by the particle must satisfy outgoing boundary conditions which can obtained by
superpositions of the VPWs with VPWs including a second type of partial wave 
$\mathbf{M}_{n,m}^{(2)}$ and $\mathbf{N}_{n,m}^{(2)}$ which are obtained by replacing the spherical Bessel functions, $j_n(x)$, in VPWs of the first type by spherical Neumann functions, $y_n(x)$. 
It will prove useful in the following to describe waves satisfying incoming boundary conditions.  The 
in-coming (-) and outgoing (+) VPWs can be expressed:
\begin{equation}
\mathbf{M}_{n,m}^{(\pm)}(k\mathbf{r})\equiv\frac{1}{2}\left( 
\mathbf{M}_{n,m}^{(1)}(k\mathbf{r})\pm i \mathbf{M}_{n,m}^{(2)}(k\mathbf{r}) \right) \label{pmVPW}
\end{equation}
and likewise for $\mathbf{N}_{n,m}^{(\pm)}$.
The scattered field can then be developed in in terms of  VPWs with outgoing 
boundary conditions again using the matrix notation:
\begin{equation}
\mathbf{E}_{\mathrm{scat}}(\mathbf{r)} = 
 [\mathbf{M}_{n,m}^{(+)}(k\mathbf{r}), \mathbf{N}_{n,m}^{(+)}(k\mathbf{r})]
\binom{f_{n,m}^{(h)}}{f_{n,m}^{(e)}}%
\end{equation}

In the multipole formulation, the $T$ \emph{matrix} by definition expresses the linear relationship between
the coefficients of the excitation field, $\mathbf{E}_{\mathrm{exc}}$ with the multipolar coefficients of the
scattered field:
\begin{equation}
\binom{f^{(h)}}{f^{(e)}}\equiv T\binom{e^{(h)}}{e^{(e)}} \label{Tdef}
\end{equation}
It is important to keep in mind in what follows that the fields $\mathbf{E}_{\mathrm{exc}}$ and 
$\mathbf{E}_{\mathrm{scat}}$ are abstractions, and  the only field which is physically present 
is the \emph{total} field, 
$\mathbf{E}_{\mathrm{tot}}=\mathbf{E}_{\mathrm{exc}}+\mathbf{E}_{\mathrm{scat}}$.

The $S$ matrix approach adopts an alternative decomposition of the total fields in terms of 
$\mathbf{E}_{\mathrm{in}}$ and $\mathbf{E}_{\mathrm{out}}$ which respectively satisfy 
incoming and outgoing boundary conditions, i.e. 
$\mathbf{E}_{\mathrm{tot}}=\mathbf{E}_{\mathrm{in}}+\mathbf{E}_{\mathrm{out}}$  with:
\begin{align}
\begin{split}
\mathbf{E}_{\mathrm{in}}(\mathbf{r)}&=[\mathbf{M}_{n,m}^{(-)}(k\mathbf{r}),
\mathbf{N}_{n,m}^{(-)}(k\mathbf{r})]\binom{a_{n,m}^{(h,-)}}{a_{n,m}^{(e,-)}} \\
\mathbf{E}_{\mathrm{out}}(\mathbf{r)}&=[\mathbf{M}_{n,m}^{(+)}
(k\mathbf{r}),\mathbf{N}_{n,m}^{(+)}(k\mathbf{r})]\binom{a_{n,m}^{(h,+)}}
{a_{n,m}^{(e,+)}}
\end{split}
\end{align}

The matrix $S$ then relates the  incoming field coefficients, $a_{n,m}^{(h,-)}$, to the outgoing 
field coefficients, $a_{n,m}^{(h,-)}$:%
\begin{equation}
\binom{a_{n,m}^{(h,+)}}{a_{n,m}^{(e,+)}}\equiv S\binom{a_{n,m}^{(h,-)}}%
{a_{n,m}^{(e,-)}} \label{Sdef}
\end{equation}
The relationship between the $S$ and $T$ matrices can be obtained algebraically by invoking the definitions, of 
Eq.(\ref{pmVPW}) and the fact that both matrices describe the same total electric field 
$\mathbf{E}_{\mathrm{tot}}$ to find:
\begin{equation}
S=I+2T
\end{equation}
where $I$ is the identity matrix. Flux conservation requires that the for a lossless  
scatterer, the unitarity of the $S$ matrix, {\it i.e. } leads directly to the energy conservation condition of 
the $T$ matrix:
\begin{equation}
T+T^{\dag}=-2T^{\dag}T
\end{equation}
which has also been called a Ward identity or ``optical theorem''.

Finally, the reaction matrix, $K$, relates the total field outside the scatterer as a superposition of regular fields, $\mathbf{E}_{\mathrm{reg}}$, developed
in terms of VPWs of the first type, and singular fields, $\mathbf{E}_{\mathrm{sing}}$, developed
in terms of the second, {\it i.e.} Neumann type:
\begin{align}
\begin{split}
\mathbf{E}_{\mathrm{reg}}(\mathbf{r)}&=[\mathbf{M}_{n,m}^{(1)}(k\mathbf{r}),
\mathbf{N}_{n,m}^{(1)}(k\mathbf{r})]\binom{r_{n,m}^{(h)}}{r_{n,m}^{(e)}}\\
\mathbf{E}_{\mathrm{sing}}(\mathbf{r)}&=[\mathbf{M}_{n,m}^{(2)}
(k\mathbf{r}),\mathbf{N}_{n,m}^{(2)}(k\mathbf{r})]\binom{d_{n,m}^{(h)}}
{d_{n,m}^{(e)}}
\end{split}
\end{align}

The $K$ matrix relates the coefficients of these two field descriptions:
\begin{equation}
\binom{d_{n,m}^{(h)}}{d_{n,m}^{(e)}}\equiv K\binom{r_{n,m}^{(h)}}{r_{n,m}^{(e)}}
\label{Kdef}
\end{equation}
For the $K$ matrix, the lossless condition is that $K$ is Hermitian, $K=K^{\dag}$.
The relations between the $K$ and the $S$ and $T$ matrices can again be obtained 
from their definitions of Eqs.(\ref{Tdef}), (\ref{Sdef}), and Eqs.(\ref{Kdef}) 
and taking into account the different VPW types:
\begin{subequations}
\begin{align}
K  & =i(S-I)(I+S)^{-1} \   \Leftrightarrow \  S=(I-iK)(I+iK)^{-1} \label{Cayley}\\
K  &  =iT(I+T)^{-1} \   \Leftrightarrow \  T^{-1}=iK^{-1}-I  \label{TK}
\end{align}
\end{subequations}
The relations between the $K$ and $S$ matrices in Eq.(\ref{Cayley}) is known as a Cayley
transformation, while Eq.(\ref{TK}) will be employed in the next section to obtain
finite size expansions of small particles. 

For spherical particles, one has analytic expressions for all the $S$, $T$ and $K$ in the
context of Mie theory where their elements are diagonal in the multipole basis,
and depend only on the total angular momentum number, $n$:
\begin{align}
\begin{split}
T_{n}^{(e)}  &  =-\frac{j_{n}(kR)}{h_{n}^{(+)}(kR)} \frac{\overline
{\varepsilon}_{s}\varphi_{n}(kR)-\varphi_{n}(k_{s}R)}{\overline{\varepsilon}_{s}
\varphi_{n}^{(+)}(kR)-\varphi_{n}(k_{s}R)} \\
S_{n}^{(e)}  &  =-\frac{h_{n}^{(-)}(kR)}{h_{n}^{(+)}(kR)}
\frac{\overline{\varepsilon}_{s}\varphi_{n}^{(-)}(kR)-\varphi_{n}(k_{s}R)}
{\overline{\varepsilon}_{s} \varphi_{n}^{(+)}(kR)-\varphi_{n}(k_{s}R)} \\
K_{n}^{(e)}  &  = - \frac{j_{n}(kR)}{y_{n}(kR)}
\frac{\overline{\varepsilon}_{s}\varphi_{n}(kR)-\varphi_{n}(k_{s}R)}
{\overline{\varepsilon}_{s}\varphi_{n}^{(2)}(kR)-\varphi_{n}(k_{s}R)}
\end{split}
\label{STK}
\end{align}
where $\overline{\varepsilon}_{s}\equiv \frac{\varepsilon_{s}}{\varepsilon_{b}}$. The 
formulas for magnetic mode response functions, $S_{n}^{(h)}$, $T_{n}^{(h)}$, $K_{n}^{(h)}$ are
obtained simply by replacing $\overline{\varepsilon}_{s}$ in the above formulas with
magnetic permeability contrast $\overline{\mu}_{s}=\frac{\mu_{s}}{\mu_{b}}$ which are the
permittivity and permeability contrasts.  
The functions, 
$\varphi_{n}$, $\varphi_{n}^{(2)}$, and $\varphi_{n}^{(\pm)}$, in eqs.(\ref{STK}) are modified logarithmic 
derivatives of the Ricatti spherical Bessel functions, and are defined as follows:
\begin{align}
\begin{split}
\varphi_{n}(z)&\equiv\varphi_{n}^{(1)}(z)\equiv  \frac{[zj_{n}(z)]^{\prime}}{j_{n}(z)}\\
\varphi_{n}^{(2)}(z)&\equiv \frac{[zy_{n}(z)]^{\prime}}{y_{n}(z)} \quad 
\varphi_{n}^{(\pm)}(z)\equiv \frac{[zh_{n}^{(\pm)}(z)]^{\prime}}
{h_{n}^{(\pm)}(z)}
\end{split}
\label{phidef}
\end{align}
with $j_{n}$ and $y_{n}$ respectively denoting $n^{\rm th}$ order spherical Bessel and Neumann
functions, and $h_{n}^{(\pm)}=j_{n} \pm i y_{n}$ the incoming $(-)$ and outgoing $(+)$ spherical Hankel functions.

\begin{widetext}

\section{Expansions for arbitrary multipole order}
\label{6ord}

We give below the development up to up to 6th order in $kR$ of the inverse reaction matrix. In most of the predictions and simulations of this work, fourth order expansion in $kR$ suffice, but the 6$^{th}$ order sometimes proved useful to test convergence or to achieve additional accuracy.
\begin{subequations}
\begin{align}
[K_{n}^{(e)}]^{-1}  & \simeq -\frac{(2n-1)!!(2n+1)!!}
{(n+1)(\overline{\varepsilon}_{s}-1)x^{2n+1}}\left(  (n\overline{\varepsilon}_{s}+n+1)
+\frac{(2n+1)((n-2)\overline{\varepsilon}_{s}+n+1)}{(2n-1)(2n+3)}x^{2} + x^{4} C^{(e)}_{4}+x^{6} C^{(e)}_{6}  \right)  \label{K1dev_elec}   \\
[K_{n}^{(h)}]^{-1} &  \simeq -\frac{\left(  2n+1\right)  
\left(2n+3\right)  \left(  2n-1\right)  !!\left( 2n+1\right)  !!}{\left(
\overline{\varepsilon}_{s}-1\right)  x^{2n+3}}\left(  1+\frac{(2n-2\overline
{\varepsilon}_{s}+3)}{(2n+1)(2n+5)}x^{2}+ x^{4} C^{(h)}_{4}+x^{6} C^{(h)}_{6}   \right)  \label{K1dev_mag}
\end{align}
\label{Kdev}
\end{subequations}
where the fourth and sixth order coefficients, respectively $C_{4}$ and $C_{6}$, are given by:  
\begin{subequations}
\begin{align}
C^{(e)}_{4} & = (2n+1) \frac{ (n+3)(n+1)^{2}+(n-4)(n+3)(n+1)
\overline{\varepsilon}_{s}-(2n-3)\overline{\varepsilon}_{s}^{2}}
{(n+1)(2n-3)(2n+3)^{2}(2n+5)} \\
C^{(h)}_{4} & = \frac{ (n+4)(2n+3)^{2}-4(n+4)(2n+3)\overline{\varepsilon}_{s}-(2n-1)
\overline{\varepsilon}_{s}^{2}  }
{(2n-1)(2n+3)(2n+5)^{2}(2n+7)} \\
C^{(e)}_{6} &= (2 n+1) \frac{ (n+1)\left(2 n^2+15 n+30\right)\left[ (n+1)  
+(n-6) \overline{\varepsilon}_{s} \right] -3 (2 n-5) \overline{\varepsilon}_{s}^2 \left[ (2 n+9)  +2  \overline{\varepsilon}_{s} \right] }
{3 (n+1)
   (2 n-5) (2 n+3)^3 (2 n+5) (2 n+7) } \\
C^{(h)}_{6} &= \frac{ (2 n+3) \left(2 n^2+19 n+47\right) \left[ (2 n+3)
   -6\overline{\varepsilon}_{s} \right]  
   -3(2 n-3)\overline{\varepsilon}^2 \left[ (2 n+11)  + 2 \overline{\varepsilon}\right]}{3 (2 n-3) (2 n+3)
   (2 n+5)^3 (2 n+7) (2 n+9)}
\end{align}
\end{subequations}
and the double factorial operator $!!$ is defined  such that:
\begin{equation}
n!!=\prod_{k=0}^{m} \left(n-2k\right) = n(n-2)(n-4)\ldots
\end{equation}
where $m={\rm Int}\left[ (n+1)/2 \right]-1 $ with $0!!=1$; or in terms of ordinary factorials via the 
relations $(2n-1)!!=\frac{(2n)!}{2^n n!}$ and $(2n)!!=2^n n!$ for $n=0,1,2,\ldots $.

\section{Exact formulas for near-field enhancements}
\label{Exact_near}
 Formally exact expressions for the field enhancement factors, Eq.(\ref{Ienh}) can be written: 
\begin{align}
\begin{split}
\left< I_{\mathrm{enh}}^{(e)}\right> &  = \sum_{n=1}^{\infty} \left\{ \widetilde{g}_{n}^{(1,e)} 
(\eta ) \left\vert T_{n}^{\left( h\right)} \right\vert ^{2}
  + \widetilde{g}_{n}^{(2,e)} (\eta ) \left\vert T_{n}^{\left( e\right)  }\right\vert ^{2} \right\}  \\
\left< I_{\mathrm{enh}}^{(h)}\right> & = \sum_{n=1}^{\infty} \left\{ \widetilde{g}_{n}^{(1,h)} (kr ) 
\left\vert T_{n}^{\left(  e\right)}\right\vert^{2}
 + \widetilde{g}_{n}^{(2,h)} (kr) \left\vert  T_{n}^{(h)}\right\vert ^{2} \right\} 
 \label{Intexact}
 \end{split}
\end{align}
 We remark that Eq.(\ref{Intexact}) is valid numerically only if the infinite multipole summation is cutoff to a value  $n_{\rm max}>kr$ with $r$ being the distance from the center of the particle.
This differs from approximate formula of Eq.(\ref{Ienh}) where the multipole summation can be
be stopped at the usual Mie cutoff condition of $n_{\rm max}>kR$.

The enhancement functions of Eq.(\ref{Intexact}) are written:
  \begin{align}
  \begin{split}
\widetilde{g}_{n}^{(1,e)}(\eta)& = \frac{ 2n+1}{2} \left\vert \frac{j_{n}(\eta)}{T_{n}^{\left( h\right)}} 
 + h_{n}^{(+)} (\eta)
  \right\vert ^{2} \\
   \widetilde{g}_{n}^{(2,e)} (\eta)  &=  \frac{ n+1}{2}  \left\vert \frac{j_{n-1}(\eta)}{T_{n}^{\left( e\right)}} +h_{n-1}^{(+)}(\eta) \right\vert^{2} +\frac{n}{2}\left\vert \frac{ j_{n+1}(\eta)}{T_{n}^{(e)}} + 
  h_{n+1}^{(+)} (\eta)\right\vert^{2}  \\
\widetilde{g}_{n}^{(1,h)}(\eta)& = \frac{ 2n+1}{2} \left\vert \frac{j_{n}(\eta)}{T_{n}^{(e)}} + h_{n}^{(+)} (\eta)
  \right\vert ^{2} \\
 \widetilde{g}_{n}^{(2,h)} (\eta)  &=  \frac{ n+1}{2}  \left\vert \frac{j_{n-1}(\eta)}
 {T_{n}^{( h)}} + h_{n-1}^{(+)}(\eta) \right\vert^{2} +\frac{n}{2}\left\vert 
 \frac{j_{n+1}(\eta)}{T_{n}^{( h)}}  +h_{n+1}^{(+)} (\eta)\right\vert^{2}
 \end{split}
 \label{exactnear}
 \end{align}

\end{widetext}


%

\end{document}